\documentclass[useAMS,usenatbib]{mnras}

\usepackage{amsmath}
\usepackage{amsfonts}
\usepackage{amssymb}
\usepackage{graphicx}
\usepackage[normalem]{ulem}
\usepackage{RefJ}

\usepackage{hyperref}
\usepackage[]{natbib}

\newcommand{\dd}{\mathrm{d}}

\usepackage{booktabs} 

\let\vec\boldsymbol 

\usepackage[dvipsnames]{xcolor} 

\usepackage{array}
\newcolumntype{T}{>{\tiny}l} 
\newcolumntype{H}{>{\Huge}l} 

\title[A method to statistically characterize turbulent data with physically motivated parameters]{A method to statistically characterize turbulent data with physically motivated parameters, illustrated on a centroid velocity map}

\author[J.-B. Durrive et al.]{Jean-Baptiste Durrive$^{1,2,3}$\thanks{E-mail:jdurrive@protonmail.com}, Pierre Lesaffre$^{2}$, Tuhin Ghosh$^{4}$, \newauthor and Bruno Regaldo-Saint Blancard$^{2,5}$\\
$^{1}$ Institut de recherche en astrophysique et plan\'etologie - Universit\'e Toulouse III - Paul Sabatier, Observatoire Midi-Pyr\'en\'ees,\\
Centre National de la Recherche Scientifique, UMR5277 - France\\
$^2$ Laboratoire de physique de l'\'ENS, \'Ecole normale sup\'erieure de Paris, Centre National de la Recherche Scientifique, FR684,\\
Universit\'e Paris Diderot-Paris 7, Sorbonne Universit\'e, UMR8023 - France\\
$^3$ Centre for mathematical Plasma-Astrophysics, Celestijnenlaan 200B, 3001 Leuven, KU Leuven, Belgium\\
$^4$ School of Physical Sciences, National Institute of Science Education and Research, HBNI, Jatni 752050, Odisha, India\\
$^5$ Observatoire de Paris, PSL University, Sorbonne Universit\'e, LERMA, 75014 Paris, France\\}

\begin{document}

\date{Accepted --- ; Received --- ; in original form ---}

\pagerange{\pageref{firstpage}--\pageref{lastpage}} \pubyear{2020}

\maketitle

\label{firstpage}

\begin{abstract}
We investigate the potential of a recently proposed model for 3D compressible MHD turbulence \citep{ChevillardEtAl10,DurriveEtAl21_BxC_Compressible} to be used as a tool to characterize statistically 2D and 3D turbulent data. This model is parametrized by a dozen of free (intuitive, physically motivated) parameters, which control the statistics of the fields (density, velocity and magnetic fields).
The present study is a proof of concept study: (i) we restrict ourselves to the incompressible hydrodynamical part of the model, (ii) we consider as data centroid velocity maps, and (iii) we let only three of the free parameters vary (namely the correlation length, the Hurst parameter and the intermittency parameter). Within this framework, we demonstrate that, given a centroid velocity map, we can find in an automated manner (i.e. by a Markov Chain Monte Carlo analysis) values of the parameters such that the model resembles the given map, i.e. which reproduces its statistics fairly well.
Hence, thanks to this procedure, one may characterize statistically, and thus compare, various turbulent data. In other words, we show how this model may be used as a metric to compare observational or simulated data sets.
In addition, because this model is numerically particularly fast (nearly 500 times faster than the numerical simulation we use to generate our reference data) it may be used as a surrogate model.
Finally, by this process we also initiate the first systematic exploration of the parameter space of this model. Doing so, we show how the parameters impact the visual and the statistical properties of centroid velocity maps, and exhibit the correlations between the various parameters, providing new insight into the model.
\end{abstract}

\begin{keywords}
turbulence -- ISM: magnetic fields -- MHD -- methods: analytical -- methods: numerical.
\end{keywords}

\section{Introduction}

Turbulence is a ubiquitous phenomenon in the Universe. There is ample observational evidence of it occurring from small to large astrophysical scales. To name but a few environments, the solar surface, the solar wind, interstellar media and intra-cluster media, are in a turbulent state, which makes their observations particularly hard to analyze.
One of the challenges is to determine on which scales the turbulent energy is injected,
because in the astrophysical context numerous phenomena occur simultaneously. The dynamics we observe in the sky involves many processes of different physical nature (gravitational, magnetic, radiative, inertial), operating over a wide range of scales \citep{ElmegreenScalo04,BrandenburgLazarian13}. For instance, stellar feedback is believed to be the main source of turbulence in the interstellar medium of our Galaxy, and active galactic nuclei appear to be playing an important role in driving intra-cluster turbulence. In addition, the information we dispose of is usually partial: often we have access only to the projections on the sky of three-dimensional fields, sometimes we lack of resolution or sensitivity, etc.
But turbulence is also difficult to analyze per se, due to its non-linear nature, which couples the scales together and behaves in a chaotic way. Its study in itself, starting from the Navier-Stokes or the magneto-hydrodynamical (MHD) equations, remains a long standing theoretical challenge, and even the Euler equations are still far from being fully understood \citep{Gibbon08}. How does the energy cascade through the scales? When and how is turbulence driven or hampered by waves and instabilities? How does turbulent dissipation occur on small scales and under which conditions it is most efficient? How do turbulent fluids self-organize to form the filamentary, ribbon-like or sheet-like structures that we routinely observe both in laboratories and in astrophysics? A vast number of questions remain yet to be answered, which are essential to understand the dynamics of astrophysical environments.

Even setting aside the importance of understanding turbulence per se, our inability to grasp and to model turbulence, magnetized or not, is problematic in practice, for interpreting astrophysical data. Facilities such as the Low-Frequency Array (LOFAR), the Square Kilometre Array (SKA), the Atacama Large Millimiter/submillimeter Array (ALMA), or space observatories such as the Advanced Telescope for High-ENergy Astrophysics (\textit{Athena}) and the James Webb Space Telescope (JWST), will soon (and some already do) enable us to probe turbulence in astrophysical (magneto-)fluids on a broad range of scales and wavelengths with unprecedented resolutions. This steady progress in instrumentation and observations provides us with data of increasingly higher quality, but consequently this data is more and more complex, and its analysis must be accompanied with adapted tools.
And indeed, in parallel, steady progress is made on the theoretical side. For astrophysically-oriented reviews on MHD turbulence see e.g. \cite{SchekochihinCowley07,BrandenburgLazarian13,Ferriere19,Rincon19,Tobias19}.

In the present work, we will focus on a specific approach to turbulence, which is complementary to numerical simulations.
Indeed, in the quest for finding explicit solutions to the incompressible Navier-Stokes equations, \cite{ChevillardEtAl10,ChevillardEtAl11,ChevillardEtAl12,ChevillardEtAl13,ChevillardHDR,PereiraEtAl16,PereiraEtAl18} built a powerful (i.e. a concise, yet fairly realistic, and numerically very efficient) analytical model for incompressible hydrodynamical turbulence.
This model was later extended to MHD in \cite{DurriveEtAl20} in the incompressible limit, soon after which compressibility effects were also considered in \cite{DurriveEtAl21_BxC_Compressible}. In \cite{DurriveEtAl20} the model was dubbed `Magnetic fields from multiplicative chaos', because the mathematical concept underlying this construction is called `multiplicative chaos' in the mathematics literature, a concept first introduced by \cite{Kahane85}.
Hence, for convenience we will refer to this model as the `BxC model', standing for `Magnetic fields from multiplicative chaos'.
We do so even though in this paper, as a first step, we will restrict ourselves to the incompressible hydrodynamical part of the model only. Hopefully our study will serve as a basis for a later extension to the magnetized case, in order to reveal the full potential of the BxC model and to widen the scope of application of the tool we construct here with it.
We will provide in section \ref{sec:Preparation} a brief introduction to the BxC model, but the bottom line is that it is an analytical model for three-dimensional MHD turbulence which is parametric, in the sense that it contains a dozen of free (physically motivated) parameters. 
In \cite{ChevillardEtAl10,ChevillardEtAl11,ChevillardEtAl12,ChevillardEtAl13,ChevillardHDR,PereiraEtAl16,PereiraEtAl18} the incompressible velocity field in this model has already been confronted to both direct numerical simulations and laboratory experiments data,
with an excellent agreement \citep[see for example section 3.4 of][for a brief review]{DurriveEtAl20}, but there is no systematic study of this model so far.

The purpose of our work is to start filling this gap, by making the first exploration of the parameter space of the BxC model, by means of a Markov Chain Monte Carlo (MCMC) analysis. From a theoretical viewpoint, this provides new insight into the model, namely on the role of each parameter and on the correlations between the various parameters. From a practical viewpoint, we aim at providing a new tool for data analysis, to help observers (as well as simulators, who deal with synthetic data) characterize statistically their turbulent data. 
The idea is that, given some turbulent data, our MCMC algorithm finds the BxC parameters which produce the BxC model with the most similar statistics as the input data. Now, since the BxC parameters are directly related to the statistical properties of the field (density, velocity or magnetic field), the best-fit parameters found by the MCMC algorithm provide a statistical characterization of this data.
In particular, a feature of turbulence which is ubiquitous in nature but that is very difficult to analyze and control, is intermittency (i.e. non-Gaussianity). In principle intermittency may be the result of extremely complex dynamics (such as a turbulent dynamo action, random shock compression, random vortex stretching, etc.), but an important advantage of BxC is that in this model intermittency is easily controlled by means of a few phenomenological (physically motivated) parameters, as detailed in section~\ref{sec:QualitativeExploration}.
Thus, the BxC model may be used as a metric between various data, by quantifying for instance how much more intermittent one data is compared to another. And this may be done in an automated way, through the procedure presented in this paper. In addition, BxC being very efficient numerically, numerous realizations may be generated in a short amount of time, such that the above best-fit BxC model can then also be used as a surrogate model.

In this paper, we present a simple example of the above procedure. The data that we aim at mimicking with the BxC model consists of a centroid velocity map, constructed from a three-dimensional velocity field that we generate using a numerical simulation, to simulate some observational data. For simplicity, and to make the presentation clearer, we vary only three parameters of the BxC model.
%
%
The paper is organized as follows. In section \ref{sec:Preparation}, we present the BxC model, and explain how we prepare the synthetic data that will constitute the reference data to which the BxC model will be confronted. In section \ref{sec:Method}, first we introduce the statistical tools that we use to characterize our maps, second we give an intuitive presentation of how the BxC centroid velocity maps vary with the BxC parameters, which helps get a feeling of how the automated parameter space exploration (i.e. the MCMC analysis) that we next introduce works. Finally, in section \ref{sec:Results}, we display the results of this MCMC analysis, showing to which extent the BxC model does resemble the given reference data, both in 2D and in 3D.

\section{Presentation of the reference data and of the BxC model}
\label{sec:Preparation}

In this section, first we detail how we prepare a synthetic centroid velocity map, that will constitute our reference data. Second, we present the BxC model and describe the parameters that we will later adjust to make the BxC model resemble the reference data.

\subsection{The reference data}
\label{sec:Data}

In astrophysical observations \citep[e.g.][]{LisEtAl96,HilyBlantEtAl08} the gas velocity is measured using the Doppler effect of spectral lines of known rest frequency emitted by the gas.
Provided the spectral resolution is good enough, spectral lines can be characterized by their moments in velocity space, and in particular the first moment, namely the line centroid
\begin{equation}
CV^{\text{obs}}(x,y)=\frac{\int v_z T(v_z) \dd v_z}{\int T(v_z) \dd v_z},
\label{def:CV_obs}
\end{equation}
where $(x,y)$ corresponds to the coordinates on the sky, the $z$ axis being chosen along the line of sight, $v_z$ is the velocity component along the line of sight, and $T(v_z)$ is the line profile, whose half-power width is proportional to the velocity dispersion of the gas. 
Relation \eqref{def:CV_obs} is the definition of centroid velocities in observational studies. Instead, in numerical studies, the centroid velocity is synthesized as the integral of a velocity component along the corresponding line of sight
\begin{equation}
CV(x,y)=\frac{1}{L_d} \int_0^{L_d} v_z(x,y,z) \dd z,
\label{def:CV}
\end{equation}
where we take $L_d$ to be the length of the periodic domain over which we perform the simulation.
Then, to better correspond to the quantities which are actually observed in emission, the centroid velocity field given by~\eqref{def:CV} can be computed by weighting the cells of the numerical grid by the local dissipation rate in the integral along the line of sight \citep[e.g.][]{Momferatos15}.
However, studying observational centroid velocity maps is out of the scope of this paper.
Therefore, for our purpose here, definition \eqref{def:CV} will be sufficient.
We produce a synthetic centroid velocity map from~\eqref{def:CV}, with a three-dimensional turbulent velocity field that we generated by running a numerical simulation presented below. We hereafter call this map the reference map, and it is shown on the top left panel of figure~\ref{fig:BxC_VS_ANK}, figure in which we gather the final results of the paper.

The numerical simulation code we used is the ANK code developed by \cite{Momferatos15}, which aims at modeling interstellar turbulence. The name of this code is the acronym of Andrei Nikolaievich Kolmogorov, in honor of the Russian mathematician and major contributor to the theory of turbulence.
It is written in the Fortran 95 programming language, implemented using the spectral method of \cite{OrszagPatterson72}, fully de-aliased by use of the phase-shift method of \cite{PattersonOrszag71}, and uses polyhedral truncation \cite{CanutoEtAl88}.
For parallelization it employs a hybrid approach, using a combination of distributed and shared memory parallelism to efficiently utilize cluster architectures of multi-core processors. Distributed parallelism is achieved using the Message Passing Interface, while shared memory parallelism is achieved using OpenMP. The code relies on the FFTW 2 free software library for the computation of Fast Fourier Transforms.
Thanks to these efforts to fasten the code, using 40 CPUs, a $256^3$ velocity field is obtained in about 4 hours, and a $512^3$ velocity field is obtained in about 10 days. We will come back to this timing in section~\ref{sec:Comparison} when comparing it to the performance of the BxC model. The resolution in our investigation will be restricted to $N=256$ points in each dimensions.
In the present work we only use the incompressible hydrodynamics part of the code, while it is is capable of simulating a wider range of dynamics. Comparing the magnetic fields from ANK to those from the BxC model is left for future work. 

\subsection{The BxC model}

The model we consider was first introduced in \cite{ChevillardEtAl10} in the hydrodynamical case, and was later generalized to MHD in \cite{DurriveEtAl20}. Here we use the notations from this more recent paper. As its name suggests, BxC is constructed for MHD, but in the present work, we will use only the velocity field of this model, which has the following explicit expression.

The starting point of the BxC model is Biot-Savart's law, which relates the velocity field $\vec{v}$ to the vorticity field $\vec{\omega}$ according to the convolution
\begin{equation}
\vec{v}(\vec{x}) = \frac{1}{4 \pi} \int_{\mathbb{R}^3} \frac{\vec{\omega} \times \vec{r}}{r^3} \dd V,
\label{BiotSavart_Hydro}
\end{equation}
with the shorthand notations
\begin{equation}
\vec{r} \equiv \vec{x}-\vec{y} \hspace{0.5cm} \text{and} \hspace{0.5cm} r \equiv |\vec{r}|,
\label{shortHandNotations}
\end{equation}
the integration being performed with respect to the $\vec{y}$ variable. Then, this expression is modified in four ways \citep[see][for more explanations]{DurriveEtAl20}: (i)~the vorticity field $\vec{\omega}$ is replaced by an intermittent random field, $\widetilde{\vec{\omega}}$, as detailed below, which transforms the expression \eqref{BiotSavart_Hydro} into a stochastic integral, accounting for the chaotic nature of turbulence, (ii)~the integration region is limited to a finite domain ($\mathcal{R}_v$ below), rather than the full $\mathbb{R}^3$ space, which introduces a large-scale correlation lengthscale ($L_v$ below), (iii)~the singular kernel $\vec{r}/r^3$ of this convolution is regularized by introducing a small-scale cut-off parameter $\epsilon_v$ (i.e. replacing in the denominator the $r^2$ by $r^2+\epsilon_v^2$, such that it does not vanish anymore), which introduces the dissipation scale, and (iv)~the power of the power-law kernel is generalized to some (a priori) arbitrary value $h_v$, called the Hurst parameter, which physically determines how `wild' the turbulence is, and geometrically determines how rough the field looks like (it is directly related to the fractal dimension). \cite{DurriveEtAl21_BxC_Compressible} introduced the expression `turbulization' to call the collection of these modifications. Hence, with this turbulization procedure, the velocity field \eqref{BiotSavart_Hydro} becomes the following modified Biot-Savart law
\begin{equation}
\vec{v} = \frac{1}{4 \pi} \int_{\mathcal{R}_v} \frac{\widetilde{\vec{\omega}} \times \vec{r}}{(r^2+\epsilon_v^2)^{h_v}} \dd V.
\label{v_tilde}
\end{equation}
The expression for the random vorticity field, $\widetilde{\vec{\omega}}$, is inspired from the theory of multiplicative chaos \citep[][]{Kahane85,RhodesVargas14}. Let $\widetilde{\vec{\omega}}_g$  be a Gaussian white noise vector (the subscript $g$ indicates a random field with Gaussian statistics), then $\widetilde{\vec{\omega}}$ is given by
\begin{equation}
\widetilde{\vec{\omega}} = e^{\tau_\omega \widetilde{\vec{\mathcal{D}}}_g} \widetilde{\vec{\omega}}_g,
\label{omega_tilde}
\end{equation}
with the randomized strain-rate matrix
\begin{equation}
\widetilde{\vec{\mathcal{D}}}_g = \frac{3}{8 \pi} \int_{\mathcal{R}_{\omega}} \hspace{-0.2cm} \frac{(\vec{r} \times \widetilde{\vec{\omega}}_g) \ \! \vec{r} + \vec{r} \ \! (\vec{r} \times \widetilde{\vec{\omega}}_g)}{(r^2+\epsilon_{\omega}^2)^{h_{\omega}}} \dd V.
\label{Dg_tilde}
\end{equation}
This matrix $\widetilde{\vec{\mathcal{D}}}_g$ also underwent a turbulization procedure, so that it contains similar parameters as in \eqref{v_tilde}, namely $\mathcal{R}_{\omega}$ which denotes the integration region, as well as $h_\omega$ and $\epsilon_\omega$ which are the Hurst parameter and dissipation scale, respectively, associated to vorticity. We invite the reader to see \cite{DurriveEtAl20} for a pedagogical justification of these expressions, which otherwise seem unintuitive, while they really are not. Indeed, the physical interpretation of \eqref{omega_tilde}, together with \eqref{Dg_tilde}, boils down to the simple fact that vortices are being stretched randomly in the fluid, and this stretching occurs during a timescale $\tau_\omega$. Since stretching is responsible for the development of non-Gaussian features, the $\tau_\omega$ parameter is called the intermittency parameter. As a simple check, notice that for $\tau_\omega = 0$, the expression~\eqref{omega_tilde} becomes $\widetilde{\vec{\omega}} = \widetilde{\vec{\omega}}_g$, i.e. in this case the vorticity field is not intermittent (it is Gaussian, and consequently, so is the velocity field), because no stretching occurred.

The above introduction to the BxC model has been very brief because in the present study we do not need the details of the construction of this model. The only thing that we will focus on, is the dependency on the parameters. 
To examine this in the clearest way, let us avoid needless complications by limiting ourselves to the simplest integration regions in the two above integrals, \eqref{v_tilde} and \eqref{Dg_tilde}: we will consider $\mathcal{R}_{v}$ and $\mathcal{R}_{\omega}$ to be balls of radius $L_v$ and $L_\omega$, respectively. We choose letter $L$ for these radii because they correspond to large-scale cut-offs, since they control the distance beyond which points in space are uncorrelated. Physically they thus model the injection scale, or integral scale, of the turbulence.

Thus, with the expression \eqref{v_tilde} for the velocity field, we get that the centroid velocity field, given by \eqref{def:CV}, can now be seen as some function, say $F$, of the various parameters of the model, i.e. we have
\begin{equation}
CV(x,y)=F(x,y;L_v,L_\omega,h_v,h_\omega,\tau_\omega,\epsilon_v,\epsilon_\omega).
\label{CV_dependence_long}
\end{equation}
Each of these parameters is a degree of freedom, useful to fit some given data. However, for simplicity, here we will fix most of them, to illustrate our method with only three free parameters. Firstly, let us choose the values of the dissipation scales, $\epsilon_v$ and $\epsilon_\omega$, to be equal to one another, and we fix them to correspond to that implemented in the ANK simulation, namely
\begin{equation}
\epsilon_v = \epsilon_\omega = \frac{2}{N},
\end{equation}
where $N$ is the number of collocation points. Taking these $\epsilon$ parameters as small as possible has the advantage of yielding an as large as possible inertial range.
Secondly, we fix $L_\omega$ to be equal to the value of the velocity field large-scale cut-off $L_v$, the latter being one of the three parameters that we left free. For convenience, hereafter we will denote this lengthscale simply by $L$, i.e. we will call the `correlation lengthscale'
\begin{equation}
L \equiv L_v = L_\omega.
\end{equation}
Thirdly, in the framework of multiplicative chaos, to ensure long range correlations, the components of the Gaussian random field which is being exponentiated (i.e. here $\widetilde{\vec{\mathcal{D}}}_g$) are usually taken to be correlated logarithmically in space. In our case, this translates into taking \citep{ChevillardEtAl10,DurriveEtAl20}
\begin{equation}
h_\omega = \frac{7}{4}.
\end{equation}
Finally, the two remaining parameters $h_v$ and $\tau_\omega$ are left free. For convenience, in the following we will get rid of their subscript, i.e. we will call the `Hurst parameter'
\begin{equation}
h \equiv h_v,
\end{equation}
the Hurst parameter of the Biot-Savart law \eqref{v_tilde}, and the `intermittency parameter'
\begin{equation}
\tau \equiv \tau_\omega,
\end{equation}
the parameter controlling the stretching of the vorticity, given by \eqref{omega_tilde}.

Altogether, with the above choices, the centroid velocity field depends only on three free parameters: $L$, $h$, and $\tau$. In other words, the relation \eqref{CV_dependence_long} becomes
\begin{equation}
CV(x,y)=F(x,y;L,h,\tau).
\label{CV_dependence}
\end{equation}
The aim of the following is to find the set of BxC parameters $(L,h,\tau)$ which results in BxC velocity fields that are as close as possible to the reference velocity field produced with the ANK simulation (reference map). We detail in the next section the method we use to find these best-fit parameters.

\section{Method for comparing the reference data to the BxC model}
\label{sec:Method}

Our basis for comparing the centroid velocity reference map to the centroid velocity maps built with the BxC model, is a set of statistics, namely the power spectrum, the spectrum of exponents and the flatness, that we now introduce.

\subsection{Statistical properties of the centroid velocity maps}
\label{sec:Statistics}

The first statistical element that we use to characterize our maps is the power spectrum. It gives information on the relative contribution of the various scales to the power of the map, and in turbulent media it follows in general a power law at intermediate wavenumbers (i.e. in the inertial range) with a large-scale cut-off, corresponding to the energy injection scale, and a small-scale cut-off, corresponding to the dissipation scale. We compute the power spectra by binning the squared amplitudes of Fourier modes with respect to the modulus of the corresponding wave number $k$. We use a regular binning in $k$ and the estimations of the power spectra are computed as the means for each bin.
In all the figures of the paper, the curves of the power spectra are presented using a light color, over which we superimpose a fit of the power spectrum presented using the same color but darker.

Since turbulent fields are in general not Gaussian fields, the power spectrum cannot fully characterize a fluid state. Hence, in addition we consider the increments of the fields and their moments, i.e. the most common tools of diagnosis in turbulence studies revealing the existence of intermittent corrections.
Specifically, let us define the velocity increment of lag $\vec{\ell}$ as the quantity
\begin{equation}
\delta_{\vec{\ell}} \vec{v}(\vec{x}) \equiv \vec{v}(\vec{x}+\vec{\ell}) - \vec{v}(\vec{x}).
\label{def:increment}
\end{equation}
Longitudinal and transverse increments are the projections of \eqref{def:increment} respectively along and perpendicular to the direction of $\vec{\ell}$.
A first classical way to reveal intermittency is to compare the probability density functions (PDFs) of the increments of the considered field to those of a Gaussian field. Indeed, in intermittent fields, the PDFs of longitudinal and transverse increments undergo a continuous deformation as the norm of the lag is decreased, from a Gaussian shape at large lags towards large tails at small lags (examples of this can be found in figures~\ref{fig:Varying_L}, \ref{fig:Varying_h}, \ref{fig:Varying_tau} and \ref{fig:BxC_VS_ANK}). These are typical signatures of intermittency, and are often called `non-Gaussian wings'. Concretely speaking, here we compute the PDFs of the increments for lags $\ell = 2, 3, 4, 5, 6, 7, 8, 12, 16, 24, 32, 48$ (the data are statistically isotropic, so only the norm $\ell$ matters). In our maps, PDFs for $\ell > 48$ are very close to Gaussians, meaning that the correlation lengthscale of our velocity fields are never greater than about a quarter of the simulation box size (of 256 pixels).

\begin{figure*}
\centering
\begin{minipage}{.5\textwidth}
  \centering
  \includegraphics[width=.9\linewidth]{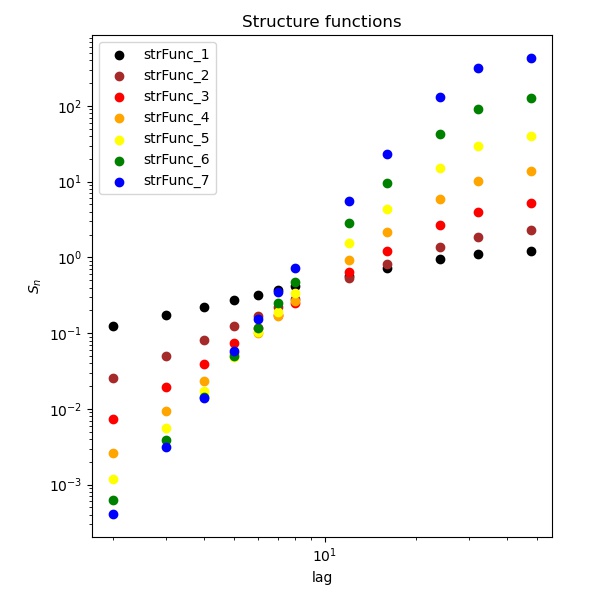}
\end{minipage}%
\begin{minipage}{.5\textwidth}
  \centering
  \includegraphics[width=.95\linewidth]{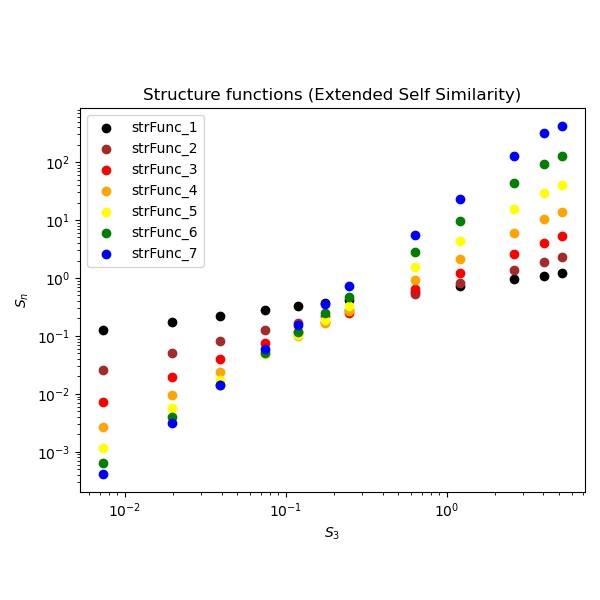}
\end{minipage}
\caption{Plots of the first seven structure functions of a centroid velocity map constructed from a realization of the BxC model. On the left we plot the structure functions with respect to the lag $\ell$, as in the original definition \eqref{def:Sn}, while on the right we use the ESS, i.e. we plot them with respect to the third order structure function, $S_3$. Comparing the two plots, it appears that the points are more aligned with the ESS, especially at large lags, meaning that the structure functions are power laws on a wider range in the ESS representation. Throughout the paper, we construct our spectra of exponents by fitting the structure functions from plots such as the right one here.}
\label{fig:ESS}
\end{figure*}

A second common way to identify intermittency in isotropic turbulence studies is to analyze the power-law behavior of structure functions in the inertial range. We use here as definition of the $n^\text{th}$ order structure function the $n^\text{th}$ moment of the absolute value of velocity increments
\begin{equation}
S_n(\ell) \equiv \langle |\delta_\ell \vec{v}|^n \rangle,
\label{def:Sn}
\end{equation}
where brackets $\langle \rangle$ indicate the expectation value \citep{Frisch95}. In practice, we use the aforementioned PDFs of increments to compute the structure functions of a given map. In the inertial range,
\begin{equation}
S_n \propto \ell^{\zeta_n},
\end{equation}
where $\zeta_n$ is called the spectrum of exponents (in fact in the following we will normalize it with the third exponent, i.e. we will consider $\zeta_n/\zeta_3$). The dependence on $n$ of $\zeta_n$ quantifies the intermittency: the field is intermittent if and only if $\zeta_n$ depends non-linearly on $n$.
Following for instance \cite{HilyBlantEtAl08}, we will actually use the Extended Self Similarity (ESS). ESS was first introduced in \cite{BenziEtAl93}, and for our purpose here it simply consists in plotting the structure functions with respect to the third order structure function $S_3$, rather than with respect to the lag $\ell$ of the increments. As shown in figure~\ref{fig:ESS}, the advantage of the ESS is that it widens the range over which the structure functions are power laws, which is convenient for finding the exponents of these power laws.

Finally, to further characterize the non-Gaussian behavior of fluctuations, the skewness
\begin{equation}
\mathcal{S}_\ell \equiv \frac{\langle (\delta_\ell \vec{v})^3 \rangle}{[\langle (\delta_\ell \vec{v})^2 \rangle]^{3/2}},
\end{equation}
and the flatness
\begin{equation}
\mathcal{F}_\ell \equiv \frac{\langle (\delta_\ell \vec{v})^4 \rangle}{[\langle (\delta_\ell \vec{v})^2 \rangle]^2},
\end{equation}
of velocity increments \citep{Frisch95} are usually investigated. For three-dimensional fields, the skewness is an essential quantity as it characterizes the energy transfer through the scales (i.e. the cascade). However, since in the following we only compute the statistics of centroid velocity maps and therefore we compute transverse increments only, our skewness is always equal to zero
for all the lags \citep[cf.][e.g.]{ChevillardEtAl10}, so that we will not mention it further. 

\subsection{Qualitative exploration of the BxC parameter space}
\label{sec:QualitativeExploration}

Before performing an exhaustive and automated exploration of the parameter space, as we will do in section \ref{sec:MCMC}, let us first get a feeling of how the centroid velocity maps and their statistics vary as we vary the BxC parameters $(L,h,\tau)$.

As a preliminary remark, let us point out that the velocity field \eqref{v_tilde}, and therefore the centroid velocity field \eqref{CV_dependence} deduced from it, is a random field who's randomness only stems from the Gaussian white noise $\widetilde{\vec{\omega}}_g$ introduced in \eqref{omega_tilde}. In other words, $\widetilde{\vec{\omega}}_g$ is the seed of the velocity field in the BxC model. Therefore, in order to examine the effects of varying the various parameters $(L,h,\tau)$, in this section we first generate one realization of the Gaussian white noise $\widetilde{\vec{\omega}}_g$, and then all the centroid velocity maps shown are derived from it. This guarantees that the evolutions we see when varying a parameter are indeed due to the variation of this parameter, and not because we would have used a different white noise realization. In that sense, our parameter space study is well-defined and deterministicdespite the inherent randomness in the model.

\begin{figure*}
\centering
\includegraphics[width=1.\linewidth]{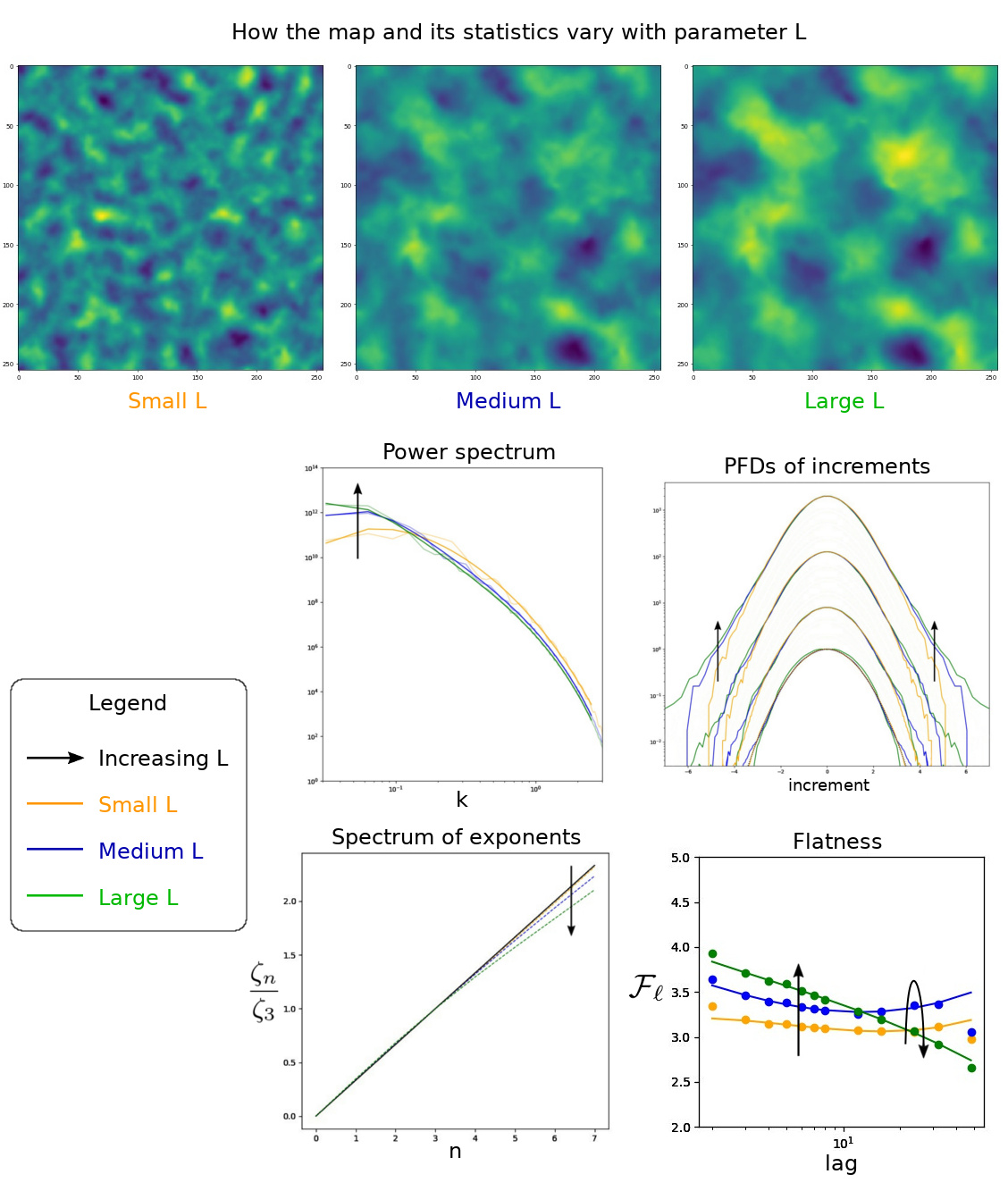}
\label{fig:test1}
\caption{Visualizations and plots showing how the appearance and the statistical properties (power spectrum, PDFs of increments for lags $\ell=2, 6, 16$ and $48$ from top to bottom, spectrum of exponents, and flatness) of the centroid velocity maps synthesized with the BxC model vary as we increase the correlation lengthscale $L$. The main role of $L$ is to control the large-scale end of the power spectrum, as indicated by the black arrow in the top left plot. This part of the power spectrum is not significantly modified otherwise, neither by $h$ nor by $\tau$ (cf. Figs~\ref{fig:Varying_h} and \ref{fig:Varying_tau}). For more discussion, see the body text.}
\label{fig:Varying_L}
\end{figure*}

Firstly, we vary the correlation lengthscale $L$, while keeping all the other parameters fixed, and the results are gathered in figure~\ref{fig:Varying_L}.
The three top panels show how the (normalized) centroid velocity maps, given by \eqref{CV_dependence}, look like for a small, an intermediate, and a large value of $L$, respectively. Clearly, as $L$ increases, the typical size of the structures appearing increases. The size of these structures is directly related to the size of the eddies of the turbulent flow. This behavior was expected since $L$ controls the size of the region over which one integrates: when in \eqref{v_tilde} one computes the velocity field at a given position $\vec{x}$, the parameter $L$ delimits how far from $\vec{x}$ positions $\vec{y}$ in the integrand will enter the integration, and therefore controls the size of the regions of correlation.
Below these three visualization panels are four plots showing some of the statistical properties of these maps. The orange, blue and green curves correspond to the small, medium and large values of $L$, respectively, and the black arrows indicate the qualitative behavior of the curves as we increase this parameter.
The top left plot corresponds to the power spectrum. Comparing the three colored curves quantifies the above discussion: the main effect of increasing $L$ is to add more power on the large scales (i.e. small wavenumbers $k$). Less expectedly, in the examples shown, increasing $L$ decreases slightly the overall shape of the power spectrum (i.e. without modifying the inertial range slope), at intermediate and smaller scales. Our interpretation is that this effect seems relatively significant here mainly because this analysis is performed with a relatively low resolution. With a higher resolution, the large-scale part of the spectrum should be more independent of the rest of the spectrum.
The top right plot shows the PDFs of increments. We normalized these histograms to unit variance and arbitrarily shifted them vertically to avoid too much curve overlapping. Also, to improve the presentation, we show only the PDFs corresponding to lags $\ell=2, 6, 16$ and $48$, from top to bottom, respectively. As a matter of interest, an example with all the computed PDFs at once is shown in figure~\ref{fig:BxC_VS_ANK}. On the bottom curve, corresponding to the largest lag $\ell=48$, we superimpose a dotted black curve corresponding to a Gaussian PDF with unit variance. By contrast, it appears that as the lag decreases, the curves deviate from this Gaussian, which is characteristic of intermittency.
These non-Gaussian wings of the PDFs become more apparent as we increase $L$, especially at small lags. On the bottom left plot, we display the spectrum of exponents. The continuous black line corresponds to the spectrum of exponents of a non-intermittent field (Kolmogorov scaling), and it appears that as $L$ increases, the departure from this linear law becomes more pronounced.
Finally, the bottom right plot corresponds to the flatness of the map as a function of the lag. It appears that the dependency of the flatness on $L$ is not straightforward: starting from small $L$ values, the flatness first uniformly increases (for all lags), until at some point it starts decreasing at large lags, as indicated by the continuous black curved arrow on the right of the plot.

\begin{figure*}
\centering
\includegraphics[width=1.\linewidth]{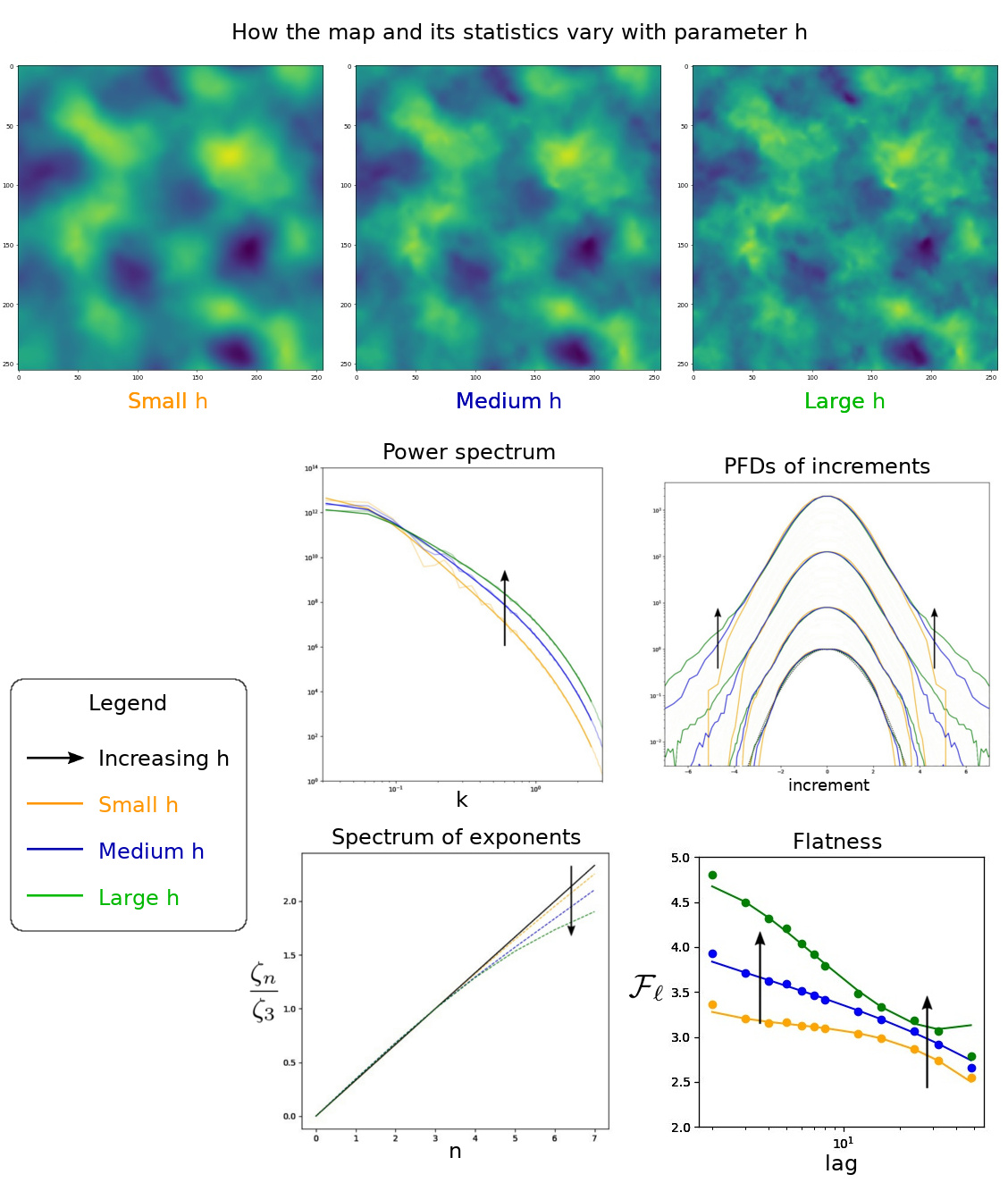}
\label{fig:test1}
\caption{Visualizations and plots showing how the appearance and the statistical properties (power spectrum, PDFs of increments for lags $\ell=2, 6, 16$ and $48$ from top to bottom, spectrum of exponents, and flatness) of the centroid velocity maps synthesized with the BxC model vary as we increase the Hurst parameter $h$. The main role of $h$ is to control the slope of the power spectrum in the inertial range, as indicated by the black double arrow in the top left plot. This slope is not significantly modified otherwise, neither by $L$ not by $\tau$ (cf. Figs~\ref{fig:Varying_L} and \ref{fig:Varying_tau}). The Hurst parameter also increases the intermittency. For more discussion, see the body text.}
\label{fig:Varying_h}
\end{figure*}

\begin{figure*}
\centering
\includegraphics[width=1.\linewidth]{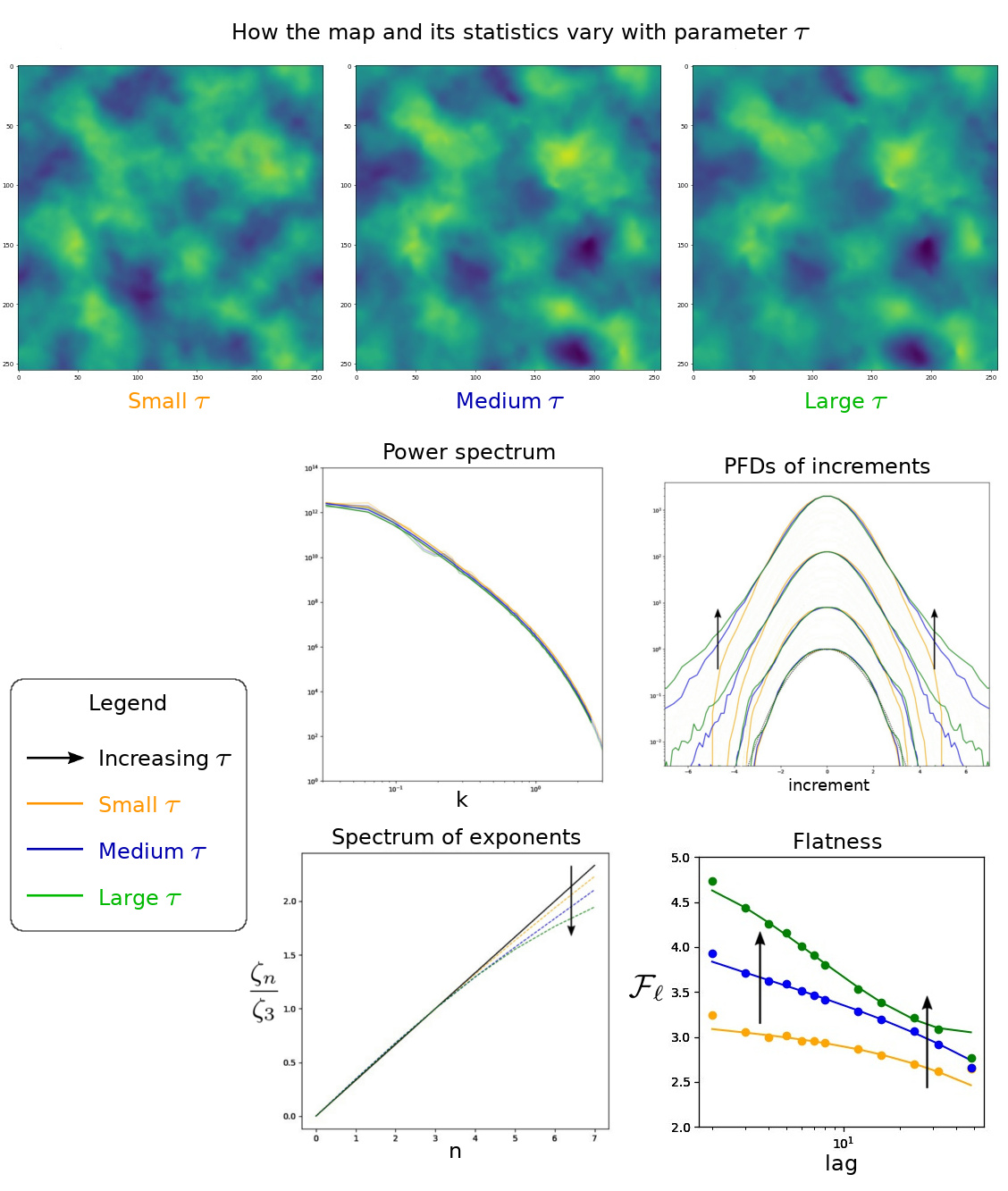}
\label{fig:test1}
\caption{Visualizations and plots showing how the appearance and the statistical properties (power spectrum, PDFs of increments for lags $\ell=2, 6, 16$ and $48$ from top to bottom, spectrum of exponents, and flatness) of the centroid velocity maps synthesized with the BxC model vary as we increase the intermittency parameter $\tau$. The main characteristic of $\tau$ is that it controls the degree of intermittency (departure from the Gaussian curve in the PDFs of increments, non-linear spectrum of exponents, and flatness different from the constant 3), without modifying the power spectrum. In that sense it deserves its name `intermittency parameter' despite the fact that $L$ and $h$ do modify it as well to some extent (cf. Figs~\ref{fig:Varying_L} and \ref{fig:Varying_h}). For more discussion, see the body text.}
\label{fig:Varying_tau}
\end{figure*}

Secondly, we vary the Hurst parameter $h$, while keeping all the other parameters fixed, and the results are gathered in figure~\ref{fig:Varying_h}. 
The three visualization panels on the top clearly indicate that $h$ controls how smooth or rough the field appears to be: for small $h$ the map looks blurry, while it contains more and more small-scale structures as $h$ grows. This illustrates the fact that Mandelbrot referred to the Hurst parameter as a measure of the `wildness' of the turbulence.
This qualitative behavior is confirmed quantitatively by the plot of the power spectrum below these panels. Indeed, the main effect of $h$ is to modify the slope of the power spectrum in the inertial range, i.e. at intermediate scales. As a matter of fact, this change is also accompanied with a slight decrease of the power on large scales, but in a less drastic manner than $L$ does in figure~\ref{fig:Varying_L}. The three other plots show that in addition, $h$ modifies the intermittency, as it makes the three signatures of non-Gaussianity appear: non-Gaussian wings in the PDFs of increments, non-linear spectrum of exponents, and a flatness departing from the non-intermittent value of 3. This fact was expected, as one facet of intermittency is related to the shape of structures, which $h$ modifies by making the turbulence become wilder.

Thirdly, we vary the intermittency parameter $\tau$, while keeping all the other parameters fixed, and the results are gathered in figure~\ref{fig:Varying_tau}. 
The first thing to notice in this figure, is that $\tau$ does indeed control the intermittency, as its name suggests, and as we expect, given the construction of the model by means of multiplicative chaos \citep[cf.][]{ChevillardEtAl10,DurriveEtAl20}. Now, one may argue that, given figures \ref{fig:Varying_L} and \ref{fig:Varying_h}, so do $L$ and $h$ to some extent. However, the interesting feature of $\tau$ is that it basically modifies \textit{only} the intermittency of the field, and deserves its name for this. Indeed, in figure~\ref{fig:Varying_tau} we choose to vary $\tau$ in a range such that its effect on the intermittency is of the same order of magnitude as the effect induced by $h$ in figure~\ref{fig:Varying_h}. And while in figure~\ref{fig:Varying_h} the power spectrum changed significantly as the intermittency varied, in figure~\ref{fig:Varying_tau} by contrast, the top left plot shows that $\tau$ does not change the power spectrum at all, while modifying the intermittency just as much (as we increase $\tau$, non-Gaussian wings of the PDFs become apparent, especially at small lags, and correspondingly the spectrum of exponents becomes more non-linear and the flatness deviates from the constant 3). Finally, to comment upon the uppermost panels, in figure~\ref{fig:Varying_tau} as we increase $\tau$, the changes in the visualizations are not as spectacular as those in figures \ref{fig:Varying_L} and \ref{fig:Varying_h}. It is so precisely because the shown range of values for $\tau$ is such that the power spectrum remains identical. Had we considered a wider range for varying $\tau$, it would have appeared that for large values its effect is, basically, to increase the contrast of the map, which can be understood by the fact that the parameter $\tau$ enters an exponential function in the velocity field formula \eqref{v_tilde}, which amplifies non-linearly the relatively large values and attenuates smaller ones. For this reason, for too large values of $\tau$ the intermittency becomes unrealistically large, and very sharp structures emerge, which already visually appear clearly unphysical.

To conclude, with this first qualitative exploration of the parameter space
we have identified the following main trends for each parameter: \textit{The parameter $L$ controls the large-scale end of the power spectrum, $h$ controls the slope of the power spectrum, and $\tau$ increases the intermittency without modifying the power spectrum}. Hence, if we were to try and fit a BxC model to some given data by hand, we would first choose $L$ to fit the large scale end of the power spectrum and then modify $h$ to fit its slope. Doing so would modify the intermittency, but we could then correct this by adjusting the $\tau$ parameter, without modifying the power spectrum.
However, in the next section we are going to use a tool which does this procedure in an automated manner, quantitatively, and tailored for generalizations in which we would vary more of the BxC free parameters.

\subsection{Automated exploration of the BxC parameter space: MCMC analysis}
\label{sec:MCMC}

In order to constrain the free parameters of the BxC model and quantify to which degree the model fits the reference map, we use the \texttt{emcee} MCMC software, written in Python by \cite{ForemanMackeyEtAl13}, and in which the Affine-Invariant sampler proposed by \cite{GoodmanWeare10} is implemented. MCMC methods are standard methods for doing Bayesian statistics, which offer insight into the correlations and degeneracies between the various parameters of the model, and are particularly suited to explore, in an automated and comprehensive way, high dimensional parameter spaces. The latter feature is essential to exploit the full potential of the BxC model. Indeed, as mentioned in equation~\eqref{CV_dependence_long}, the hydrodynamical part of the BxC model already contains seven free parameters, and it contains about twice more free parameters once magnetic fields are included \citep[cf. table 1 of][]{DurriveEtAl20}, and even a few more still can be varied once the full BxC model is considered, i.e. when compressibility is taken into account \citep{DurriveEtAl21_BxC_Compressible}. Hence, the present study, in which we limit ourselves to varying only three parameters, should serve as a basis, i.e. a first simple example
to be generalized. 

Concretely speaking, we proceed as follows.
Firstly, we build a centroid velocity map, given by expression \eqref{def:CV}, from a 3D velocity field generated with the ANK simulation (cf. section~\ref{sec:Data}). Secondly, we compute the statistical properties of this map, namely its power spectrum, its spectrum of exponents and its flatness (cf. section~\ref{sec:Statistics} for the definitions, and cf. figure~\ref{fig:BxC_VS_ANK} for the results). Thirdly, we construct a data vector $\vec{d}$, with 11 entries, as follows. We fit the inertial range of the power spectrum (indicated by the black arrow in the plot of the power spectrum of figure~\ref{fig:BxC_VS_ANK}) to a linear law, which produces two fitting parameters $(d_1,d_2)$ corresponding to the slope and the y-intercept of that linear law, respectively. We then fit the flatness with a third order polynomial, which produces three additional fitting parameters, the values of which are given to the entries $(d_3,d_4,d_5)$ of the vector~$\vec{d}$. And finally, we use six of the seven exponents of the spectrum of exponents to complete this vector data (i.e. we use all of the exponents, except the third one which is trivially equal to 1 since we normalize by $\zeta_3$). In other words, the vector components $(d_6,d_7,d_8,d_9,d_{10},d_{11})$ correspond to $(\zeta_1/\zeta_3,\zeta_2/\zeta_3,\zeta_4/\zeta_3,\zeta_5/\zeta_3,\zeta_6/\zeta_3,\zeta_7/\zeta_3)$.

To constrain the BxC parameters, we maximize the log-likelihood function ($\text{\sc{T}}$ denotes transposition)
\begin{equation}
\mathcal{L} = -\frac{1}{2} (\vec{d}-\vec{m})^{\text{\sc{T}}} \vec{C}^{-1} (\vec{d}-\vec{m}),
\end{equation}
where $\vec{d}$ is the aforementioned data vector, namely the concatenation of the parameters fitting the statistics of the reference map, and $\vec{m}$ corresponds to the model vector, namely an analogous vector to $\vec{d}$ but built from a BxC generated centroid velocity map. In other words, $\vec{m}$ is the concatenation of the parameters fitting the statistics of the BxC maps generated at the various steps of the MCMC run. The matrix $\vec{C}$ is the covariance matrix, considered diagonal.
Finally, finding a relevant guess to initialize the Markov chains, as well as finding appropriate priors, is rather straightforward thanks to the preliminary qualitative exploration of the parameter space outlined in section \ref{sec:QualitativeExploration}. Indeed, first, since $L$ basically corresponds to the size of the largest coherent structures appearing in the reference map, we can guess simply visually from the upper left panel of figure~\ref{fig:BxC_VS_ANK} that $L=0.25$ (i.e. a quarter of the box size) is a relevant order of magnitude for this parameter, and that $0.1<L<0.4$ is an appropriate prior. Second, we vary the Hurst parameter $h$ to adjust roughly the slope of the power spectrum, finding that reasonable values for $h$ are around unity, so we start with $h=0.8$, and restrict the parameter space with the constrain $0.3<h<1.2$.
Finally, we choose to initialize the intermittency parameter to $\tau=3$ and limit it to the range $0<\tau<6$, because $\tau=0$ is an obvious lower bound as it corresponds to a Gaussian field, $\tau=6$ happens to yield centroid velocity maps in which unphysically sharp structures appear (as mentioned towards the end of section~\ref{sec:QualitativeExploration}), so half of its value, $\tau=3$, is chosen as initialization since it is a compromise between having a Gaussian map ($\tau=0$) and an excessively non-Gaussian map ($\tau=6$).

\section{Comparison of the reference data to the BxC model}
\label{sec:Results}

In this section we present how our best-fit BxC model compares to the reference data.
The results are gathered in figure~\ref{fig:CornerPlot}, displaying the corner plot summarizing the MCMC analysis, and in figure~\ref{fig:BxC_VS_ANK}, with both a visual (qualitative) and a statistical (quantitative) comparison of the reference data with the best-fit BxC model.

\subsection{Best-fit BxC model}
\label{sec:Comparison}

\begin{figure*}
\centering
\includegraphics[scale=0.65]{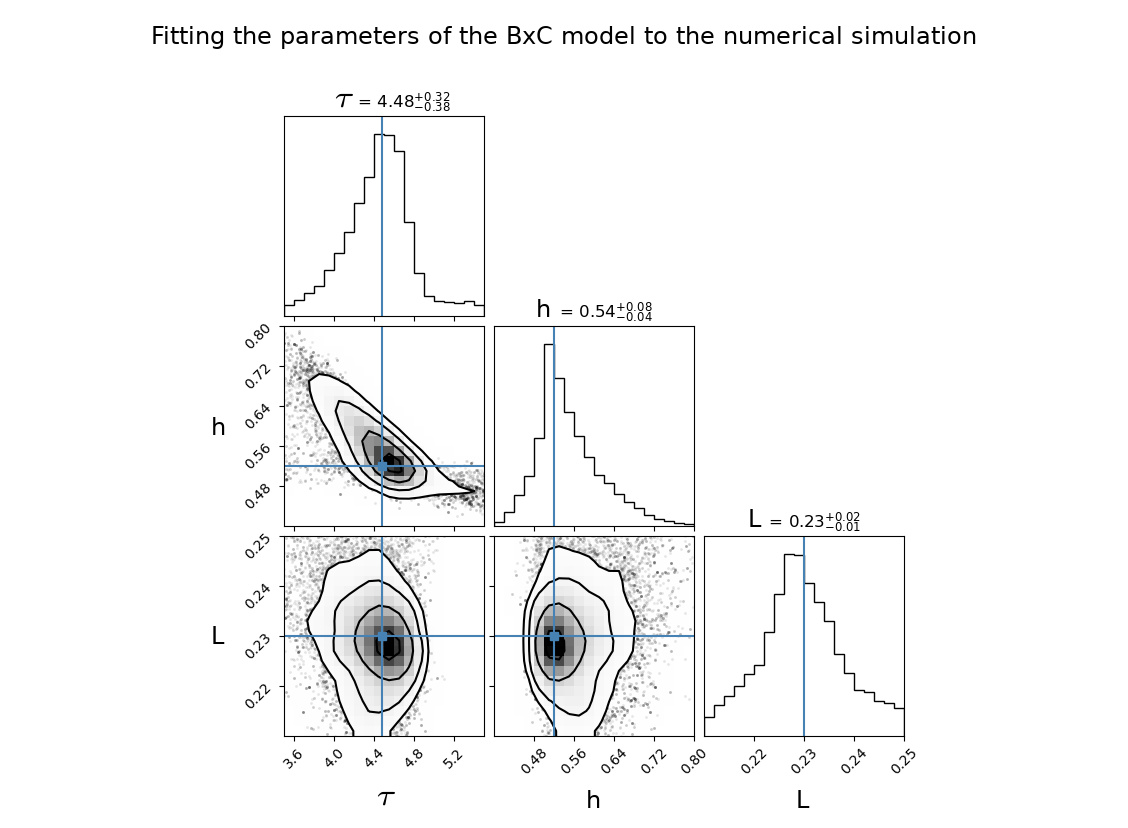}
\caption{Corner plot derived from our MCMC analysis. Panels along the diagonal are the one-dimensional histograms of the BxC parameters $(L,h,\tau)$ obtained by marginalizing over the other parameters. Panels off the diagonal are the two-dimensional projections of the posterior probability distributions for each pair of parameters. Blue lines point out the best-fit values: an example of a realization of a centroid velocity map with these best-fit parameters is shown in the top right panel of figure~\ref{fig:BxC_VS_ANK}.
}
\label{fig:CornerPlot}
\end{figure*}

At first, as a sanity check of our MCMC analysis code, we first did the following test. We generated some reference data with the BxC model itself, i.e. we generated a synthetic centroid velocity map which had some underlying known BxC parameters, say $(L_\text{test},h_\text{test},\tau_\text{test})$.
We then ran two MCMC analyses, one in which we initialized the run with large values, $(1.2 L_\text{test}, 1.2 h_\text{test}, 1.2 \tau_\text{test})$, and the other in which we initialized the run with small values, $(0.8 L_\text{test}, 0.8 h_\text{test}, 0.8 \tau_\text{test})$, and checked whether, in both cases, the algorithm did converge towards the input parameters or not. We obtained corner plots similar to the one shown in figure~\ref{fig:CornerPlot} (the latter being
detailed below), where the best-fit parameters did correspond to the parameters from which the map was built from, as it should. This confirmed that our MCMC algorithm behaved properly.

\begin{figure*}
\centering
\includegraphics[scale=0.65]{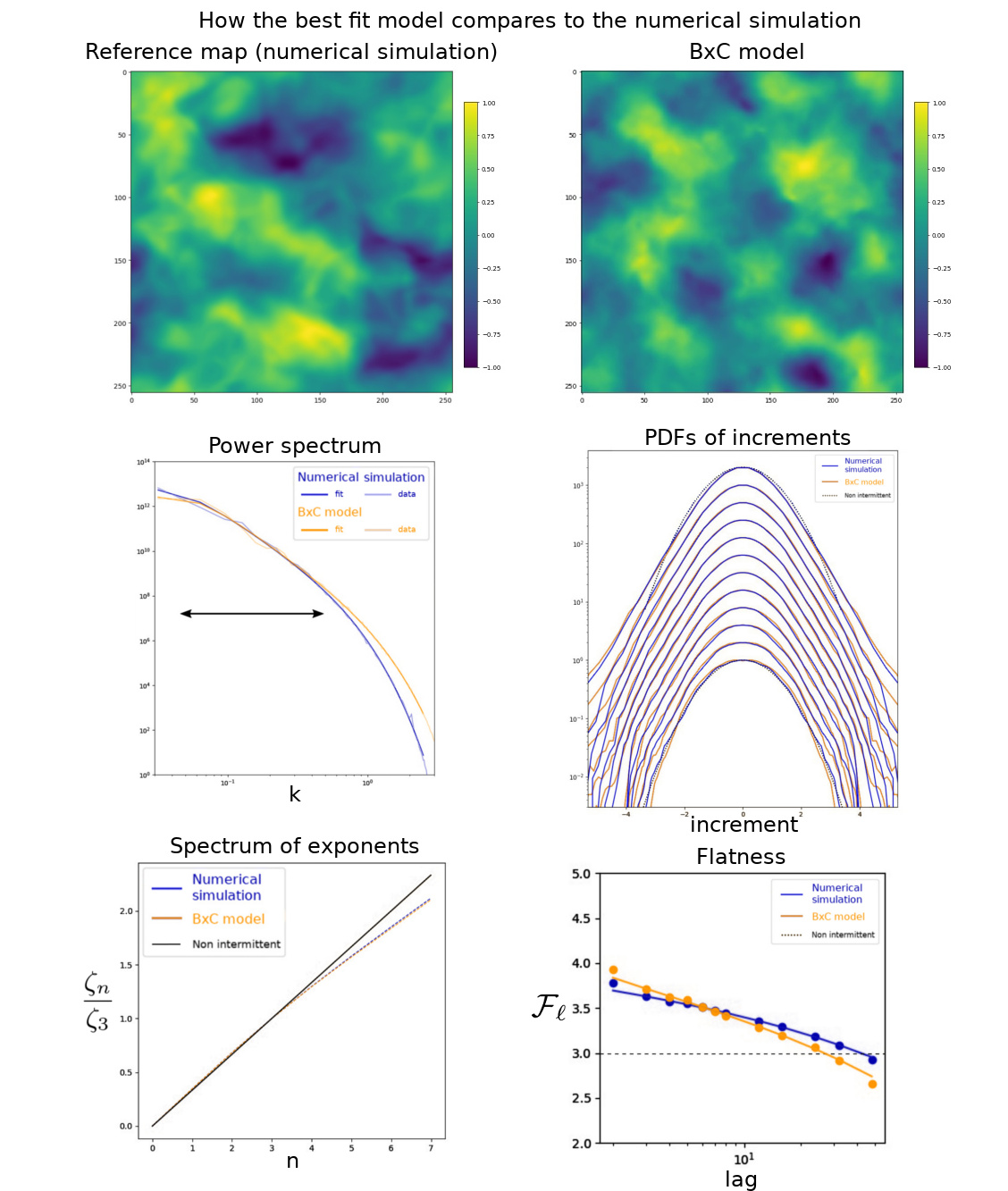}
\caption{Visual and statistical comparison of the centroid velocity maps built with the reference data and a realization of the BxC model with the best-fit parameters from figure~\ref{fig:CornerPlot} (blue lines in that figure). The black double arrow in the power spectrum indicates the inertial range, over which the fitting is done. For more discussion, see the body text.}
\label{fig:BxC_VS_ANK}
\end{figure*}

We then applied this method to the reference data from the ANK numerical simulation presented in section \ref{sec:Data}.
We used 120 walkers, with 300 steps and a burn-in phase of 100 steps.
The first point to stress is that in the corner plot of figure~\ref{fig:CornerPlot}, the one-dimensional marginalized posterior distributions (the diagonal panels) peak at specific values, i.e. the algorithm converged towards some preferred values, namely $(L,h,\tau)=(0.23,0.54,4.46)$, indicated with the blue lines in the figure. Therefore, one specific BxC model is favored by the data. In figure~\ref{fig:BxC_VS_ANK} we then show how this best-fit BxC model compares with the reference map, both visually, and in terms of statistical properties. The two maps, at the top of the figure, look similar enough to feel that indeed they possibly have similar statistics. We confirm this feeling quantitatively with the plots below the maps. The power spectra of the two data sets fit in the inertial range (admittedly, the latter, indicated by the black double arrow, is rather small here because we work with a relatively low resolution, but the inertial range of the BxC model becomes a more pronounced power law as resolution increases). As in figures~\ref{fig:Varying_L}, \ref{fig:Varying_h} and \ref{fig:Varying_tau}, the three plots neighboring the plot of the power spectrum correspond to the diagnosis tools revealing intermittency. With a fairly good agreement between reference data and BxC data, the PDFs of increments are non-Gaussian, the spectrum of exponents is non-linear and the flatness deviates from 3. In conclusion, figure~\ref{fig:BxC_VS_ANK} shows that the best-fit BxC map does mimic fairly well the reference map.

To complete the analysis, let us comment on the relevance of these numerical values.
An attractive feature of the BxC model is that its construction is intuitive, and the parameters entering the description have simple physical meanings (eddy size, vortex stretching, dissipation lengthscale, etc).
However, admittedly, a downside of having generalized the power in the power-law kernel of the Biot-Savart law \eqref{BiotSavart_Hydro} is that due to this the intermittency parameter $\tau$ does not have the physical dimension of a time for an arbitrary $h$, while in the construction it enters as the correlation timescale of the velocity gradient. 
Hence, the BxC model is meant to provide a convenient (fast and intuitive) but only effective description. We obviously do not expect it to perfectly match velocity (or magnetic) fields neither from numerical simulations nor from observations, because the formulae of the model are just approximations of the solutions of the Navier-Stokes (and MHD) equations.
Nevertheless, the whole point and the important benefit of having found numerical values to the parameters $(L,h,\tau)$ is that they caracterize in a simple manner the turbulent (hence intricate) reference map. This way, we provide a quantitative manner to compare various maps, such as those shown in figures~\ref{fig:Varying_L}, \ref{fig:Varying_h} and \ref{fig:Varying_tau}. For instance, had the map on the top left of figure~\ref{fig:Varying_h} and the one on the top right been two given maps to be compared, with the present MCMC analysis we would have been able to deduce the BxC parameters best describing each of them, and we would thus have been able to quantify how more turbulent the map on the right is compared to the one on the left. In other words, BxC may be used as a metric between data sets.

In addition to revealing best-fit values, the corner plot in figure~\ref{fig:CornerPlot} provides us with confidence intervals.
Elongated confidence intervals are the sign of correlations between parameters. The most striking correlation that we exhibit here, which appears in the two-dimensional plot in the second row of figure~\ref{fig:CornerPlot}, is the anti-correlation between the Hurst exponent and the intermittency parameter.
It is the first time that this anti-correlation is displayed in the three-dimensional BxC model. This result generalizes an earlier similar result: in \cite{ChevillardHDR} a one-dimensional ersatz of the present three-dimensional velocity model was introduced, the statistical properties of which have been computed analytically, and in particular, it was shown that in order for the model to satisfy the 4/5 law of turbulence, a linear relation between the Hurst exponent and the intermittency parameter had to be imposed.
Physically, we may interpret the existence of such a correlation from the fact that intermittency has two facets, statistical and structural \citep[e.g.][]{HilyBlantEtAl08}. 
Roughly speaking, $h$ could in a sense be referred to as some `structural intermittency' parameter, because its main effect is to modify the shape of the structures, adding more small scale structures as it increases (cf. figure~\ref{fig:Varying_h}), and it does so uniformly in the whole of space, so that it hardly modifies the volume filling of the field. By contrast, $\tau$~seems more like a `statistical intermittency' parameter, as it tends to change the volume filling of the field. Indeed, since it is the parameter controlling the exponential function in the velocity field~\eqref{v_tilde}, as $\tau$ increases the contrast in the centroid velocity map increases, making sharp regions sharper, which is not a spacially uniform effect since it acts more on some regions than others, as opposed to the uniform increase of small scale structures by~$h$.

\subsection{A strong asset of the BxC model: Timing}
\label{Timings}

Numerical simulations are the relevant tool for generating synthetic data which includes precisely a wide range of physical processes at once, resulting in fairly realistic data. On the contrary, the physics underlying the BxC model is rather elementary (streching, compression and shear), such that the description it provides is essentially effective. However, the power of the BxC model compared to numerical simulations, is the speed at which it can generate fairly realistic data. To be specific, in the present case, despite all the efforts mentioned in section \ref{sec:Data} to fasten the ANK code, with a given amount of resources (namely 40 CPUs), a $256^3$ velocity field is obtained in 4 hours and 7 minutes, and a $512^3$ velocity field is obtained in 238 hours (about 10 days). Conversely, with the BxC code that we used in this work, a $256^3$ velocity field is obtained in 34 seconds, and a $512^3$ velocity field is obtained in 3 minutes and 47 seconds. Hence, in theses cases, BxC is 460 and 3800 times faster than the numerical simulation, respectively.
It is precisely this speed which enabled us to run an MCMC analysis.

\begin{figure*}
\centering
\includegraphics[scale=0.4]{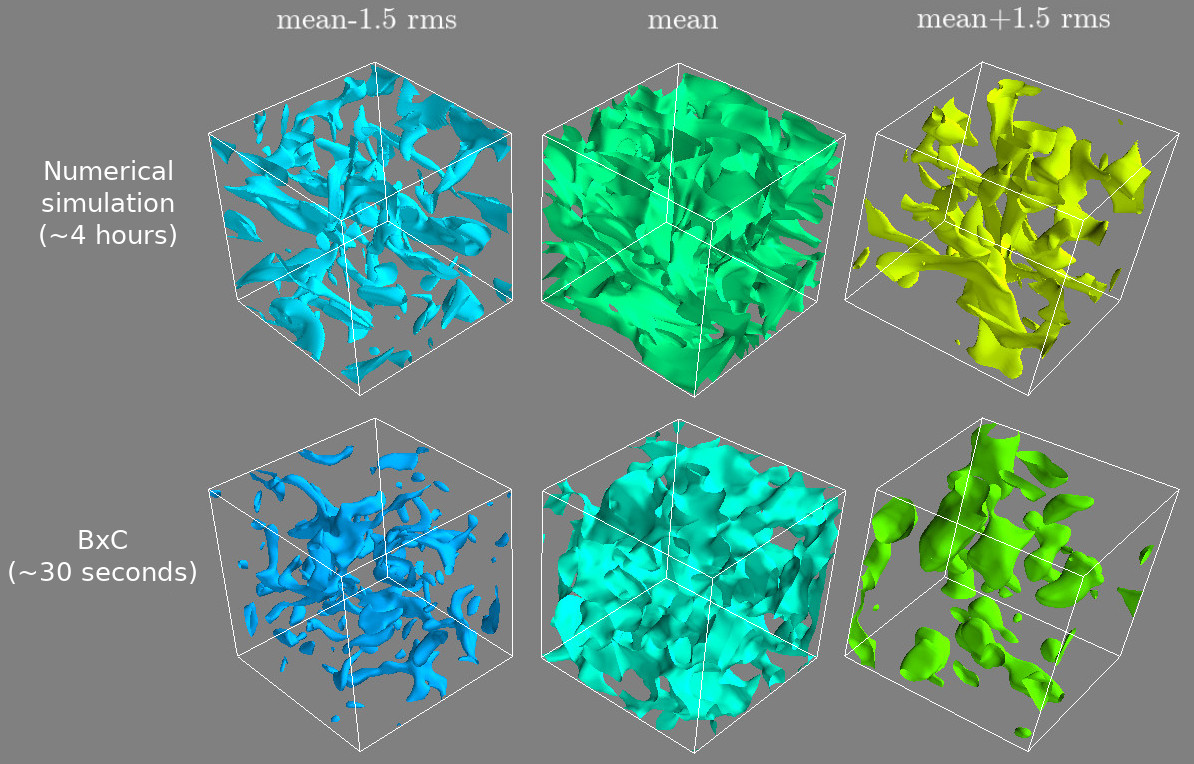}
\caption{Top: Three-dimensional visualizations of the iso-contours of the velocity field used to construct the reference map in our analysis. From left to right, the contours correspond to the mean value minus $1.5$ times the root mean square, the mean and the mean value plus $1.5$ times the root mean square, respectively. Bottom: Same iso-contours as above but with a realization of the best-fit BxC model. On the far left column we also indicate orders of magnitude of the time it took to generate these fields, using the same computing resources.}
\label{fig:3D}
\end{figure*}

\subsection{Restoring the three-dimensional picture}
\label{sec:3D}

We have identified best-fit parameters by comparing the statistics of two-dimensional data. Had we used an observational map, the analysis would have been over. However, since we in fact have three-dimensional data here, we may look if the reference velocity field and the best-fit BxC velocity field are also similar in three dimensions, and not only in terms of their projections. 
A precise comparison is out of the scope of the paper, and we limit ourselves to a visual, qualitative, comparison here to stress one particular point. On the top row of figure~\ref{fig:3D} we show three iso-contours of the velocity field that we used for the reference data, and on the bottom row we show the same contours obtained with a realization of the best-fit BxC model. The figure was produced using Mayavi, an application and library for interactive scientific data visualization and 3D plotting in Python \citep{RamachandranVaroquaux11}.
Visually, the two fields overall resemble each other fairly well. However, looking closer at the structures individually, one notices that the shape of the structures in the numerical simulation are clearly anisotropic, often almost flat, while those in the BxC model tend to be more ellipsoidal. One of the reasons for this is that in the present analysis we limited ourselves to spherically symmetric integration regions $\mathcal{R}_v$ and $\mathcal{R}_\omega$ (cf. equation \eqref{CV_dependence_long} where we introduced the parameter $L$ for simplicity). But still, this difficulty to produce filaments, ribbons, and sheets is a weakness of the BxC model that was already pointed out by \cite{DurriveEtAl20} and \cite{DurriveEtAl21_BxC_Compressible}, who provided ideas on how to improve this in future versions of the model: using anisotropic regularization norms, non-spherical integration regions, or modifying the `initial' conditions in the model by focusing on the invariants of the velocity field gradient (i.e. considerations of the (P,Q,R) space) for instance.

\section{Conclusion and prospects}

BxC is an analytical model for three-dimensional compressible MHD turbulence, which contains a dozen of free (physically motivated) parameters, which control the statistics of the fields (density, velocity and magnetic fields). It was first introduced in the incompressible hydrodynamical case by \cite{ChevillardEtAl10} and later \cite{DurriveEtAl20,DurriveEtAl21_BxC_Compressible} proposed an extension to MHD with compressibility effects. We have initiated the first systematic exploration of the parameter space of this model.
Our study is a proof of concept study (we restricted ourselves to characterizing centroid velocity maps, with the incompressible hydrodynamical case, and varied only three of the parameters), in which we have demonstrated how the BxC model may be used as a tool for analyzing turbulent data.
To do so, we have generated with a numerical simulation a synthetic centroid velocity map, and, by means of an MCMC algorithm, we found in an automated manner a BxC model which resembles the synthetic reference map, i.e. which reproduces its statistics fairly well.

Thanks to this procedure, one may characterize statistically, and then compare, various turbulent data sets. In other words, the BxC model may be used as a metric to compare observational or simulated data.
Another major asset of the BxC model is that it is numerically particularly fast, namely almost 500 times faster than the numerical simulation we used to generate our synthetic data.
This speed enabled us to perform our MCMC analysis,
but the other advantage of this performance is that the BxC model may thus be used as a surrogate model.

In addition, through this study we have exhibited how the parameters we vary impact the visual and statistical properties of the centroid velocity maps. We also showed, by means of the MCMC analysis, how the parameters depend on each other, and in particular revealed, for the first time in this three-dimensional model, the anti-correlation between the Hurst and intermittency parameters. This indicates that the linear relation between these two parameters, previously established in a one-dimensional version of the BxC model \citep{ChevillardHDR}, also arises in the full three-dimensional version.

Being meant as a tool for data analysis, the most natural next step would be to confront BxC to real data, rather than synthesized data. The above analysis may be directly applied to observations of the interstellar medium, such as those in molecular clouds with the centroid velocity maps of \cite{HilyBlantEtAl08} for example. This tool is also relevant for observations of the intra-cluster medium,
notably for the forthcoming \textit{Athena} data. For these studies it will however be essential to vary more parameters than what we did here, in particular letting the dissipation scale vary, since it is a crucial information on the underlying turbulence. Such an extension would be most easily achieved, since MCMC methods are precisely tailored for high dimensional problems.
In addition, in order to reveal the full potential of the BxC model (rather than limit ourselves to incompressible hydrodynamics as above) and widen its applications, it would be worth analyzing for example polarization maps (with $(Q,U)$ Stokes parameters reference maps), first using the incompressible magnetic fields of \cite{DurriveEtAl20} and then their extension with compressibility introduced in \cite{DurriveEtAl21_BxC_Compressible}.

It would also be worth improving the analysis itself, which may be done in several ways.
An obvious weakness of the above analysis is that we used a relatively low resolution, with $N=256$ collocation points, such that notably the inertial ranges of our velocity fields were not striking. It would thus be important to perform a similar MCMC analysis with $N=512$, which is still affordable thanks to the speed of BxC. This would tell whether some of the effects observed here are resolution dependent or not. It may indeed be so, as for example the inertial range of BxC velocity fields gets wider as the resolution increases, we expect the parameter $L$ to modify more the large-scale cut-off of the power spectrum than its slope as it does here.
We could also improve our tool of diagnosis, i.e. the statistical properties that we choose to compare the data and the model. Here we did not consider a precise constraint on the shape of the structures. We could for instance consider the invariants of the velocity gradient, i.e. the geometry of the (Q,R) plane, which is one way to quantify statistically the shape of structures \citep{Meneveau11}.

\section*{Acknowledgments}

We thank F. Boulanger, E. Falgarone and K. Ferri\`ere for fruitful discussions. This research is supported by the Agence Nationale de la Recherche (project BxB: ANR-17-CE31-0022) and the European Research Council (Advanced Grant MIST (FP7/2017-2022, No 742719)).

\bibliographystyle{mnras}
\bibliography{BxC_ANK}

\end{document}